\documentclass[
reprint,
superscriptaddress,
 amsmath,amssymb,
 aps,
]{revtex4-2}

\usepackage{graphicx}
\usepackage{dcolumn}
\usepackage{bm}

\begin{document}

\preprint{APS/123-QED}

\title{Spin Phase Continuous Modulation: A Method for the Measurement of Neutron Monochromaticity}

\author{Ryuto Fujitani}
 \affiliation{Department of Nuclear Engineering, Kyoto University, Kyoto 615-8246, Japan}
 \affiliation{Institute for Integrated Radiation and Nuclear Science (KURNS), Kyoto University, Kumatori, Osaka 590-0494, Japan}
 \email{fujitani.ryuto.85m@st.kyoto-u.ac.jp}

\author{Masahiro Hino}
 \affiliation{Institute for Integrated Radiation and Nuclear Science (KURNS), Kyoto University, Kumatori, Osaka 590-0494, Japan}

\author{Takashi Higuchi}
 \affiliation{Research Center for Nuclear Physics (RCNP), The University of Osaka, Osaka 567-0047, Japan}
 \affiliation{Institute for Integrated Radiation and Nuclear Science (KURNS), Kyoto University, Kumatori, Osaka 590-0494, Japan}


\date{\today}

\begin{abstract}
We present spin phase continuous modulation (SPCM), a new method for characterization of neutron beam properties.
By utilizing spin-precession modulation induced by two oscillating magnetic fields, the method enables precise determination of neutron velocity and monochromaticity.
We derive a theoretical formulation of the SPCM signal and experimentally validate it using measurements with a monochromatized neutron beam, 
obtaining reasonable agreement between experiment and prediction.
SPCM enables quantitative determination of neutron beam velocity and monochromaticity.
\end{abstract}

\maketitle


\section{Introduction} 
One of the most important aspects of neutron beam science, including neutron scattering, is accurately determining the velocity of a neutron beam. 
Beam performance can be evaluated using a criterion called monochromaticity, which is described as $\delta v/v$. 
Here, $v$ is the neutron velocity and $\delta v$ is the velocity spread. 
Since large scale pulsed neutron sources such as J-PARC, SNS and ESS have become operational,
measurements of monochromaticity and velocity using time-of-flight (TOF) method have become increasingly important\cite{maekawa2010first, kajimoto2019status, mason2005spallation, andersen2020instrument}.
While the TOF method can measure velocity based on the flight time and path length of neutrons, 
the instrumental limitations of beamlines cause the measured velocity to exhibit a finite spread. 
Consequently, the monochromaticity is also a finite value. 
Therefore, calibration is required for precise TOF measurements, 
which relies on scattering from well-characterized crystalline references\cite{torii2011super, shinohara2020energy, shibata2015performance, mamontov2011time}.
Although crystal-based calibration is highly precise, 
it cannot resolve spatial profiles and constrains detector position because it relies on scattered beams. 
Additionally, the obtained spectrum can only be evaluated over a limited range due to the monochromatization of the crystals.

We propose a new method, named as spin phase continuous modulation (SPCM), to measure the monochromaticity and neutron velocity without scattering or monochromatization.
SPCM uses spin phase modulation induced by two oscillating magnetic fields (modulators).
From the phase and its dispersion of modulation, the velocity and monochromaticity can be measured in high precision.
Inspired by methods developed for the precision determination of neutron velocity in fine-structure constant measurements\cite{Krueger_1995, krueger_1998final, Krueger_2000},
SPCM has been developed for the measurement of monochromaticity.
The methodology is based on polarized neutron techniques\cite{mezei1972neutron, baryshevskii1991neutron, mezei2002neutron, gahler1992neutron, ebisawa1998quantum, YAMAZAKI2002623, suzuki-fujitani}.
In this paper, we present a proof-of-principle demonstration of monochromaticity and neutron velocity measurement, along with the corresponding analysis.

\section{Formulation of Spin Phase Continuous Modulation}
A SPCM system consists of a neutron spin interferometer.
The detected neutron count $N$ in a neutron spin interferometer is given by
\begin{equation}\label{eq:NSI_pattern}
  N = \frac{N_0}{2} \left\{ 1 - \cos (\chi - \Omega - \delta) \right\},
\end{equation}
where $N_0$ is the maximum neutron count, 
$\chi$ is the phase difference between the two resonance spin flippers (RSFs),
$\delta$ is the phase shift due to the guide magnetic field applied to maintain the spin orientation
and $\Omega$ is the phase induced by a magnetic field between RSFs, corresponding to the spin precession phase.

In a SPCM system, 
two oscillating magnetic fields parallel to the quantization axis are placed between the RSFs.
Consider a case in which the modulators are separated by a distance $l$.
When a neutron passes the center of the first magnetic field $x=0$ at $t=\tau$, the phase $\Omega$ induced by the magnetic fields is given by
\begin{equation}\label{eq:OmegaTwoOsci}
  \begin{aligned}
    \Omega(\tau) & = \frac{2|\mu_n|}{\hbar v} \left\{ \int_{-\infty}^{\infty} B_0(x) \sin\!\left(\frac{2\pi f}{v}x + \phi_1 + 2\pi\tau \right) dx \right. \\
    & \left.\qquad + \int_{-\infty}^{\infty} B_0(x-l) \sin\!\left(\frac{2\pi f}{v}x + \phi_2 + 2\pi\tau \right) dx \right\}. \\
  \end{aligned}
\end{equation}
Here, $\mu_n$ is the magnetic moment of the neutron, $\hbar$ is the Dirac constant and $v$ is the neutron velocity.
$B_0 (x)$ is the magnetic field profile, $f$ is the frequency of the oscillating magnetic fields, and $\phi_1$ and $\phi_2$ are phases of the fields.
Assuming the magnetic fields are zero outside of the RSFs, the range has been extended to infinity.

Defining $k = 2\pi f /v$ and introducing the Fourier transform of the magnetic field profile in $k$-space $b(k)$, defined as
\begin{equation}\label{eq:bk}
  |b(k)|e^{i\theta(k)} = \int_{-\infty}^{\infty} B_0(x) e^{-ikx} dx,
\end{equation}
Eq.~(\ref{eq:OmegaTwoOsci}) can be rewritten as
\begin{equation}\label{eq:OmegaTwoOsci2}
  \begin{aligned}
    \Omega(\tau) & = \frac{2|\mu_n|}{\hbar v} |b(k)| \sin\!\left(\frac{\pi f}{v}l +  \frac{\phi_1 + \phi_2}{2} + 2\pi\tau + \theta(k)\right) \\
    &\qquad \qquad \qquad \times 2 \cos\!\left(\frac{\phi_1 - \phi_2}{2} - \frac{\pi f}{v}l\right).\\
  \end{aligned}
\end{equation}
Using the imaginary part of Eq.~(\ref{eq:bk}), Eq.~(\ref{eq:OmegaTwoOsci2}) is derived from Eq.~(\ref{eq:OmegaTwoOsci}).
As neutrons pass continuously, $\tau$ varies with time, and accordingly the phase $\Omega(\tau)$ also changes.
This implies that the spin precession phase is modulated by the oscillating magnetic fields, following a sinusoidal function.
Thus, the oscillating magnetic fields work as modulators.
As a result, the resulting interfernce pattern is 
\begin{equation}
  N = \frac{N_0}{2} \lim_{T\rightarrow\infty} \frac{1}{2T}\int^{T}_{-T} \left\{ 1 - \cos (\chi - \Omega(\tau) - \delta) \right\} d\tau.
\end{equation}
By defining $\theta=2\pi\tau$, the integration range can be transformed into a finite interval.
\begin{equation}\label{eq:SPCM_1}
  \begin{aligned}
    N &= \frac{N_0}{2} \frac{1}{2\pi}\int^{\pi}_{-\pi} \left\{ 1 - \cos (\chi - \Omega(\theta) - \delta) \right\} d\theta\\
    & = \frac{N_0}{2} \left\{ 1 - J_{0}\left(\frac{4|\mu_n|}{\hbar v} |b(k)| \cos\!\left(\frac{\Delta \phi}{2} - \frac{\pi f}{v}l\right)\right) \right. \\
    & \left. \qquad\qquad \times \cos (\chi - \delta) \right\}.
  \end{aligned}
\end{equation}
Here, $\Delta \phi = \phi_1 - \phi_2$ is introduced.
Eq.~(\ref{eq:SPCM_1}) shows that the modulation in spin precession phase appears as a change of the contrast in the interference pattern.
As a result, periodic contrast change depending on $\Delta\phi$ and $\pi fl/v$ can be obtained.

By setting $\chi - \delta=\pi$, the contrast can be directly obtained from the neutron count.
\begin{equation}\label{eq:SPCM}
  N = \frac{N_0}{2} \left\{ 1 + J_{0}\left(\frac{4|\mu_n|}{\hbar v} |b(k)| \cos\!\left(\frac{\Delta \phi}{2} - \frac{\pi f}{v}l\right)\right)\right\}.
\end{equation}
This is the formulation of SPCM signals.
SPCM signals can be obtained as neutron count $N$.
According to Eq.~(\ref{eq:SPCM}), the SPCM signal exhibits peaks when $\Delta \phi / 2 - \pi fl/v = \pi (n + 1/2), n \in \mathbb{Z}$ is satisfied.

For a finite neutron velocity distribution, phase dispersion appears in a SPCM signal.
As a result, the SPCM signal is smeared along the phase $\Delta \phi$.
When the neutron velocity distribution is expressed as $g(v^{\prime})$,
the SPCM signal, including the effect of velocity dispersion, can be written as
\begin{equation}\label{eq:SPCM_disp}
 \begin{aligned}
    N &= \frac{N_0}{2} \left\{ 1 + \int g(v^{\prime}) \right.\\
    & \left. \qquad\qquad  \times J_{0}\left(\frac{4|\mu_n|}{\hbar v^{\prime}} |b(k^{\prime})| \cos\!\left(\frac{\Delta \phi}{2} - \frac{\pi f}{v^{\prime}}l\right)\right) dv^{\prime} \right\}.
 \end{aligned}
\end{equation}
The neutron velocity distribution is convoluted with SPCM signal.

\section{Experimental Setup for Spin Phase Continuous Modulation}

To demonstrate the measurement of a neutron velocity and its dispersion using SPCM, 
we conducted an experiment at JRR-3 C3-1-2-2 (Multilayer Interferometer and reflectometer forNEutron2, MINE2), 
where monochromated cold neutron beam is available.

\begin{figure*}[t]
 \begin{center}
  \includegraphics[width=\textwidth]{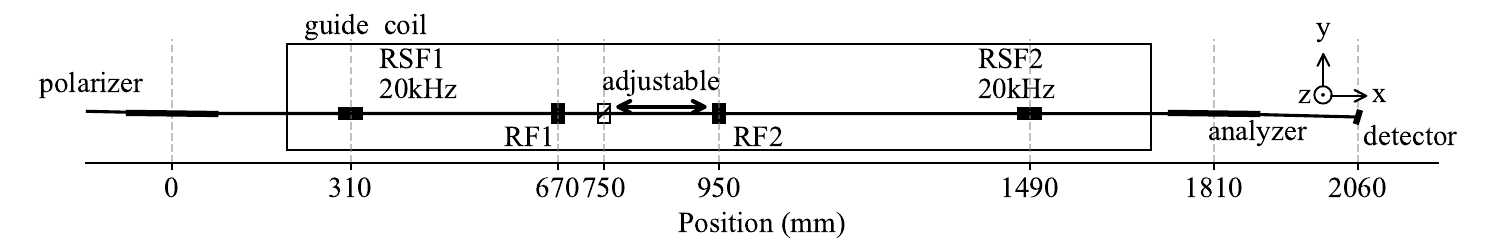}
  \caption{Experimental setup for spin phase continuous modulation (SPCM) at JRR-3 C3-1-2-2. }
  \label{fig:interferometerSetup}
 \end{center}
\end{figure*}

The experimental setup is shown in FIG.~\ref{fig:interferometerSetup}.
The beam direction is along the $x$-axis.
The SPCM setup consists, in order from the upstream, of a polarizer, RSF1, RF1, RF2, RSF2, an analyzer, and a neutron detector.
A guide magnetic field is applied throughout the entire spin interferometer system to maintain the neutron spin quantization axis.

\begin{figure}[t]
 \begin{center}
  \includegraphics[width=\linewidth]{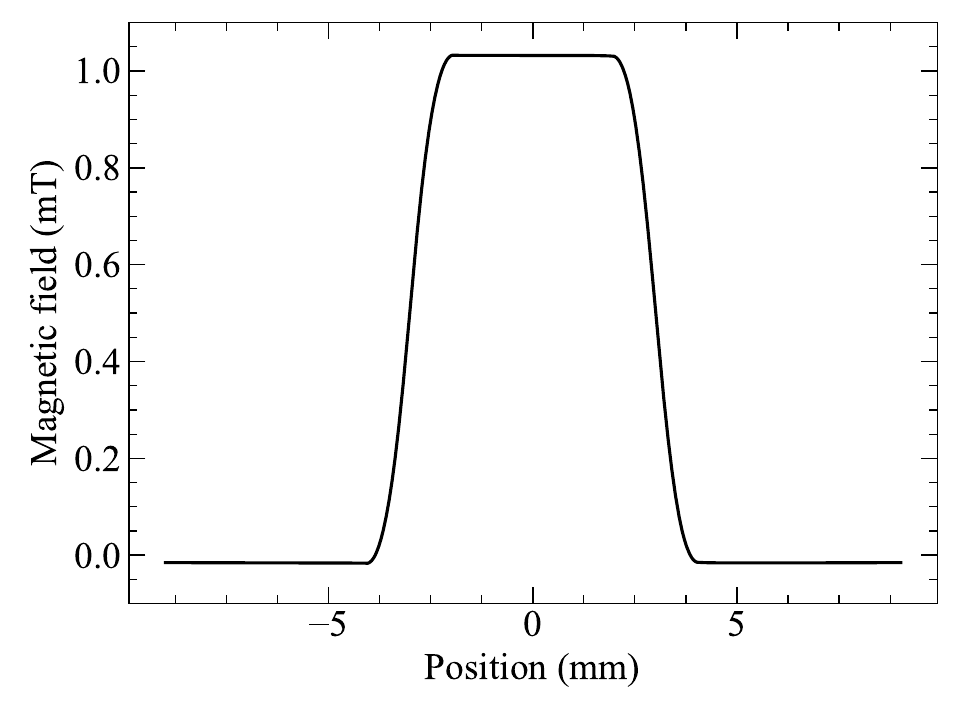}
  \caption{Magnetic field induced by the RF coil for 1~A current}
  \label{fig:Bz}
 \end{center}
\end{figure}

The polarizer and analyzer, which reflect only spin-up neutrons, are composed of magnetic multilayer mirrors.
The RSFs, which flip the neutrons spin, generate 20~kHz oscillating magnetic fields along the $x$-axis.
RF1 and RF2 generate oscillating magnetic fields along the $z$-axis to induce phase modulation in spin precession.
RF1 and RF2 are constructed by winding a 1.1mm diameter aluminum wire around a 5mm thick acrylic plate.
The coil has dimensions of 120~mm in the $z$-direction (height) and 60~mm in the $y$-direction (width) and consists of 100 turns.
The position of RF1 is fixed on the beam line.
RF2 is mounted on a linear stage with a positioning accuracy of $15~\mathrm{\mu m}$, 
allowing adjustment of the distance between RF1 and RF2 in the range from 80~mm to 280~mm.
FIG.~\ref{fig:Bz} shows a magnetic field generated at the coil center along the $x$-axis when a current of 1~A is aplied.
This field is calculated based on the Biot-Savart law.

\begin{figure}[t]
 \begin{center}
  \includegraphics[width=\linewidth]{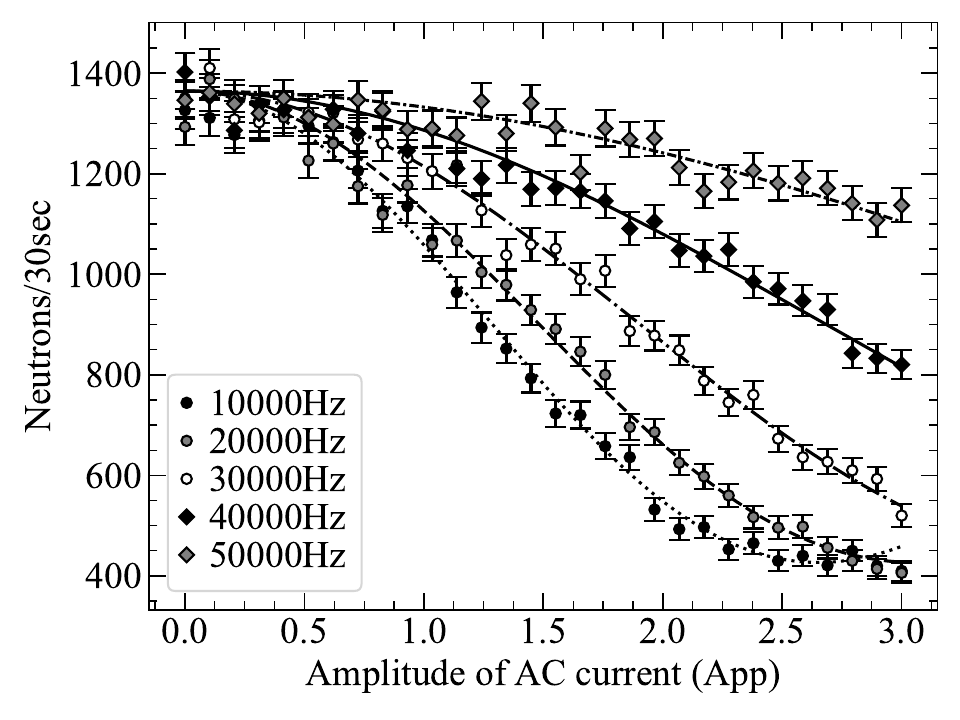}
  \caption{Neutron counts as a function of the current applied to the RF coil}
  \label{fig:RFs_bessel}
 \end{center}
\end{figure}

\begin{figure}[t]
 \begin{center}
  \includegraphics[width=\linewidth]{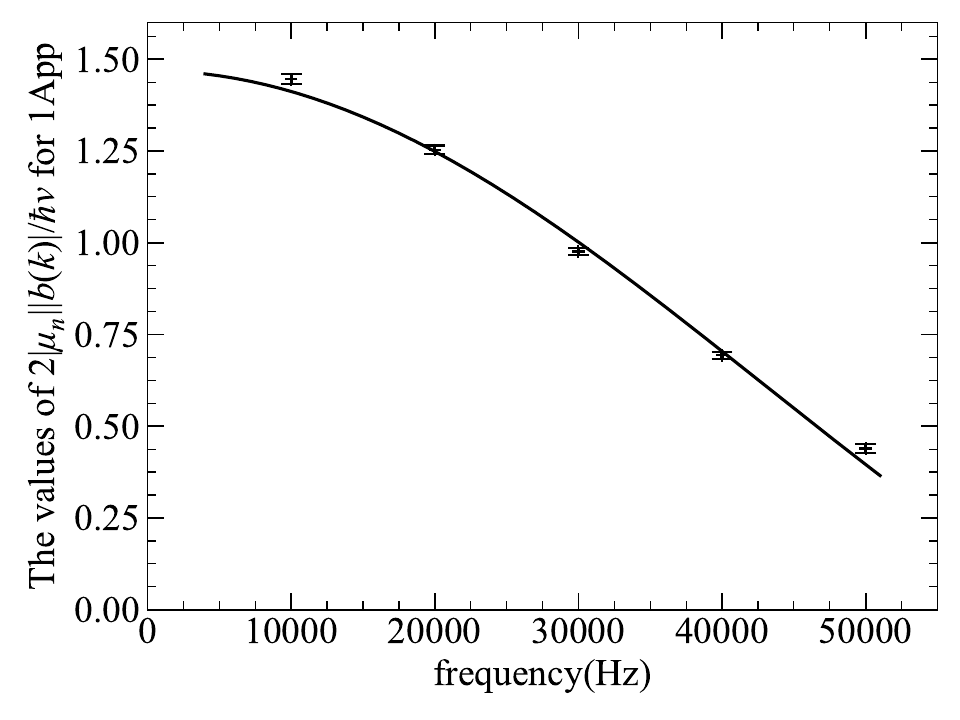}
  \caption{Value of $2|\mu_n||b(k)|/ \hbar v$ obtained from the fitting to Eq.~(\ref{eq:bk_check})}
  \label{fig:bkfit}
 \end{center}
\end{figure}

The property of the RF coils were inspected using a neutron spin interferometer by a method described in a previous paper\cite{suzuki-fujitani}.
By installing a RF coil in a neutorn spin interferometer and measuring the time-averaged neutron counts, resulting neutron counts will be
\begin{equation}\label{eq:bk_check}
  N = \frac{N_0}{2} \left\{ 1 - J_{0}\left(\frac{2|\mu_n|}{\hbar v} |b(k)| \right) \cos (\chi - \delta) \right\}.
\end{equation}
FIG.~\ref{fig:RFs_bessel} shows the neutron counts obtained by changing the current $I$ applied to the RF coil while the interferometer phase is fixed to satisfy $\chi - \delta = \pi$.
The obtained neutron counts are fitted with the function $N = \frac{N_0}{2} (1+CJ_0(\alpha I))$ to determine the value of $2|\mu_n||b(k)|/ \hbar v$.
Here, $N_0$ and $C$ are the maximum neutron count and the contrast of the neutron spin interferometer, respectively, obtained from independent calibration measurements.
$\alpha$ is a fitting parameter.
The obtained values of $2|\mu_n||b(k)|/ \hbar v$ for a current of 1~App are shown in FIG.~\ref{fig:bkfit}.
If the magnetic field is approximated as a rectangular profile with length $L$ and strength $B$,
the values of $2|\mu_n||b(k)|/ \hbar v$ can be expressed as
\begin{equation}\label{eq:bk_rect}
  \frac{2|\mu_n|}{\hbar v} |b(k)| = \frac{2|\mu_n|}{\hbar} \frac{B}{\pi f L} \left| \sin \left(\frac{\pi f L}{v} \right) \right|.
\end{equation}
Thus, the data are fitted to the function $|A_1 \sin(A_2 f)/f|$, where $A_1$ and $A_2$ are fitting parameters.

In the SPCM experiments, an AC current of 3~App was applied to RF1 and RF2.
RF1 and RF2 are connected to the amplifier with opposite polarities.
The frequency has been changed among 10~kHz, 20~kHz and 30~kHz.
The phase of RF1 was fixed at 0~deg, while that of RF2 was scanned from 0~deg to 720~deg.
The distance between RF1 and RF2 was changed among 80~mm, 130~mm, 180~mm, 230~mm and 280~mm.

\section{Results and Discussion}

\begin{figure*}[t]
 \begin{center}
  \includegraphics[width=\textwidth]{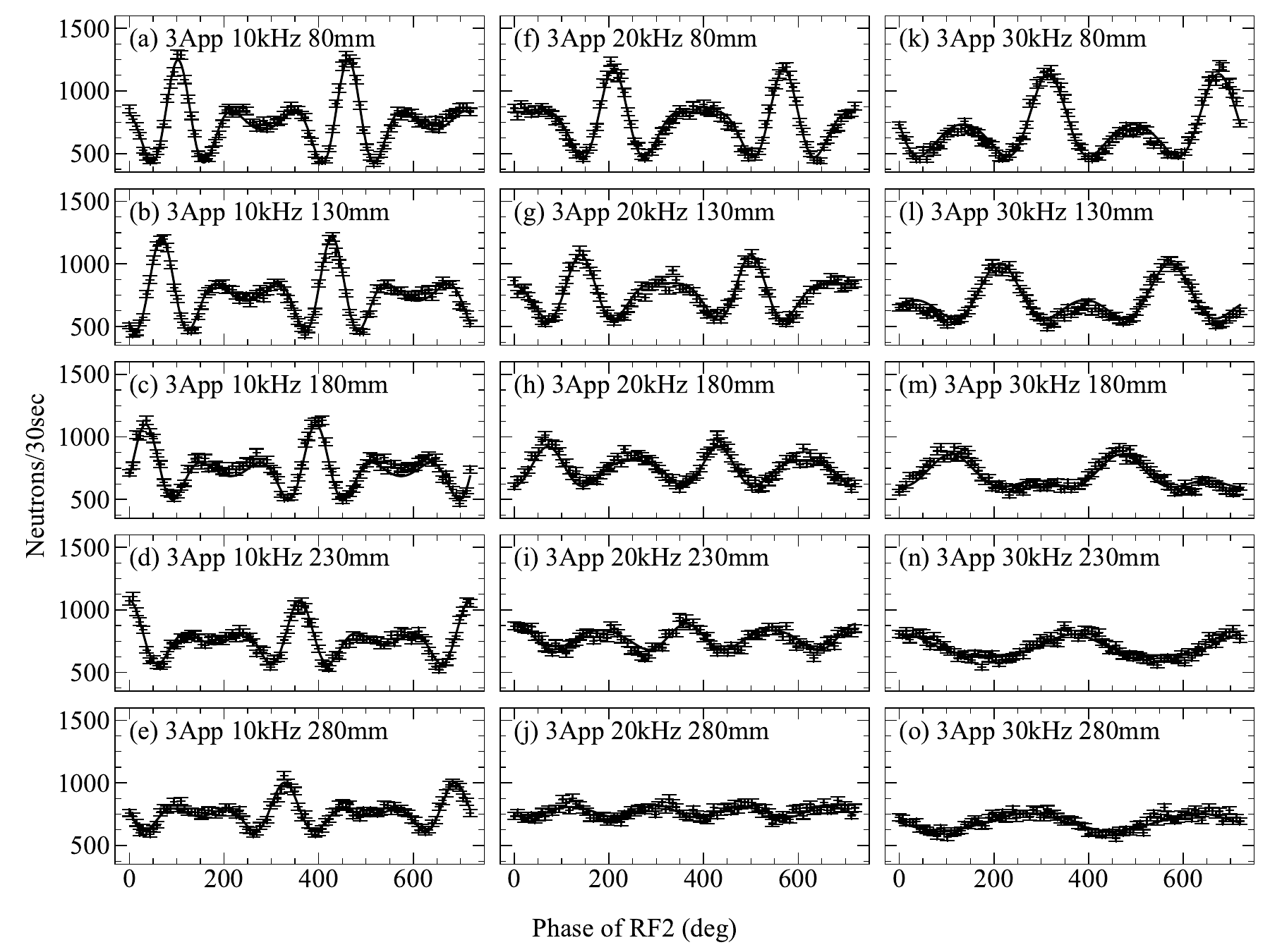}
  \caption{SPCM signals and fitting results}
  \label{fig:SPCM}
 \end{center}
\end{figure*}

The results of the SPCM experiment are shown in FIG.~\ref{fig:SPCM}.
The horizontal axis represents the phase of RF2 $\phi_2$, 
the vertical axis shows the neutron count $N$.
The data are fitted to
\begin{equation}\label{eq:SPCM_fit}
    \begin{aligned}
        N &= \frac{N_0}{2} \left\{ 1 + C \int \frac{1}{\sqrt{2\pi \sigma^2}}e^{-\frac{(p-P_0)^2} {2\sigma^2}} \right. \\
        & \left. \qquad \qquad \times J_{0}\left(A_p \cos\!\left(\frac{-\phi_2-p}{2}\right)\right) dp \right\}.
    \end{aligned}
\end{equation}
In Eq.~(\ref{eq:SPCM_disp}), the neutron velocity distribution shown in Eq.~(\ref{eq:SPCM_fit}) is considered to be a normal distribution of the phase in a SPCM signal.
$N_0$ and $C$ are maximum neutron count and the contrast of the neutron spin interferometer,
which were obtained from calibration measurements conducted prior to each experiment.
$A_p$ is the parameter determined from FIG.~\ref{fig:bkfit}, representing the value of $2|\mu_n||b(k)|/ \hbar v$.
$\sigma$ and $P_0$ represent the phase dispersion and the phase offset of the SPCM signal, respectively, and are used as fitting parameters.

\begin{figure}[t]
 \begin{center}
  \includegraphics[width=\linewidth]{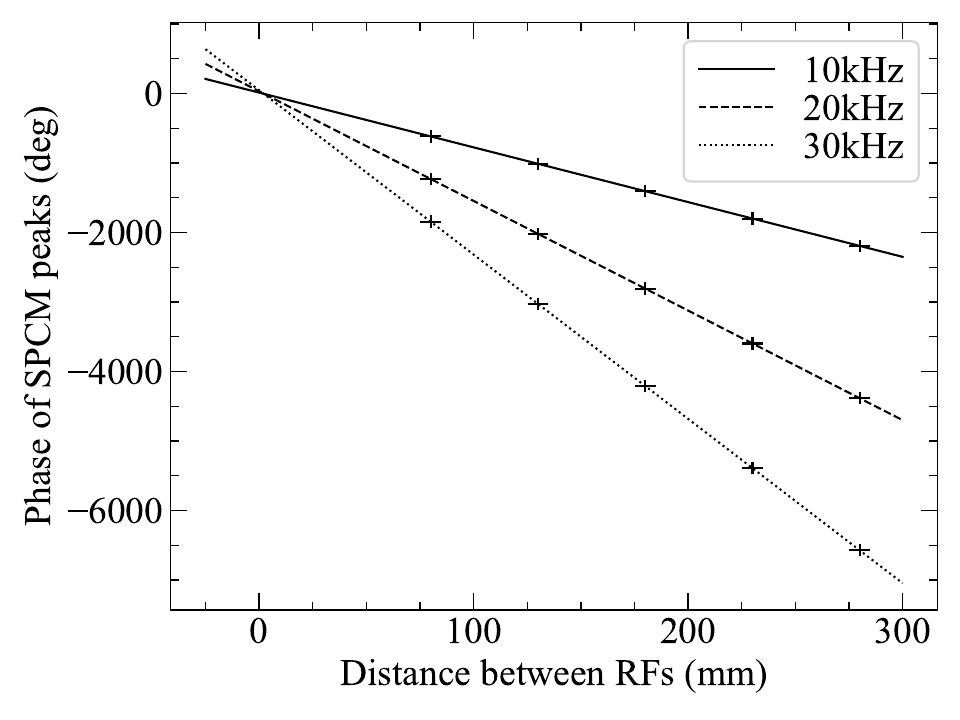}
  \caption{The peak positions of SPCM signals}
  \label{fig:SPCM_phaseShift}
 \end{center}
\end{figure}

The obtained phase $P_0$ of the SPCM signal is shown in FIG.~\ref{fig:SPCM_phaseShift}.
The horizontal axis represents the distance between RF1 and RF2, 
and the vertical axis represents the phase of the SPCM peak position, $\pi - P_0$.
The lines correspond to linear fits for each frequency.
The values of $P_0$ are chosen to satisfy $|P_0 - 2\pi fl/v_{rough}| \le \pi$.
Here, $v_{rough}$ represents approximate neutron velocity.
We adopt $v_{rough} = 450~\mathrm{m/s}$ based on the beamline specifications.
If the neutron velocity is completely unknown, 
SPCM measurements with lower frequencies and shorter distances between RF1 and RF2 are sufficient.

The slopes of the lines are expected to follow $2\pi f/v$.
Thus, by fitting all data with a function $P_0 = - 2\pi f (l-x_0)/v_0$, treating $v_0$ and $x_0$ as fitting parameters, 
we obtained $v_0 = 456.438 \pm 0.090 ~ \mathrm{m/s}$ and $x_0 = 1.955 \pm 0.025 ~ \mathrm{mm}$.
Since the RF coils are connected with opposite polarities, the peaks are expected to appear at zero phase of RF2 around $l = 0$ and $x_0 = 0$ conditions.
While the relative displacement between the RF coils is controlled by a linear stage with an accuracy of $15~\mathrm{\mu m}$, 
the initial distance was measured using a ruler.
Therefore, a discrepancy of several millimeters between the actual and nominal distances, as reflected in $x_0$, is reasonable.
Even if the absolute distance between the RF coils is not accurately known, the neutron velocity can be determined as long as the relative distance changes are accurate.
Furthermore, the absolute distance can be corrected by evaluating $x_0$.

\begin{figure}[t]
 \begin{center}
  \includegraphics[width=\linewidth]{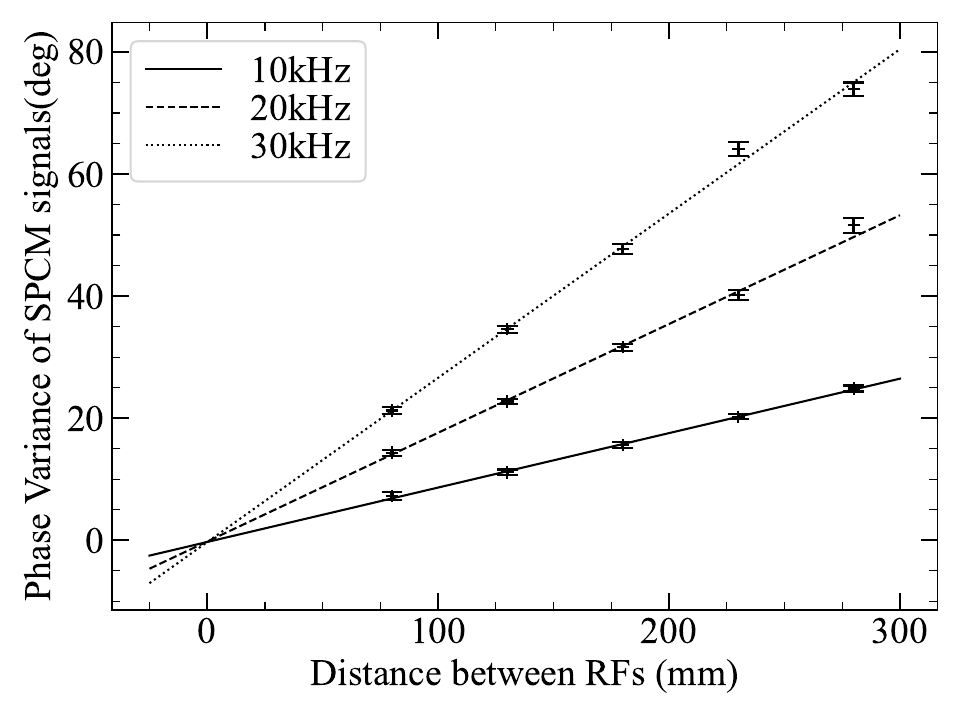}
  \caption{The phase dispersion of SPCM signals}
  \label{fig:SPCM_phaseDisp}
 \end{center}
\end{figure}

The obtained phase dispersion $\sigma$ of the SPCM signal is shown in FIG.~\ref{fig:SPCM_phaseDisp}.
The horizontal axis represents the distance between RF1 and RF2, and the vertical axis represents the phase dispersion of the SPCM signal $\sigma$.
The lines correspond to linear fits for different frequency.
Assuming that the phase dispersion follows $\sigma = 2\pi f (l-x_1)/v_0 \times dv/v_0$, where $dv$ donates a velocity dispersion,
all data are fitted with $dv$ and $x_1$ as fitting parameters.
As a result, we obtain $dv = 5.158 \pm 0.072 ~ \mathrm{m/s}$ and $x_1 = 0.912 \pm 2.256 ~ \mathrm{mm}$.
The obtained velocity dispersion corresponds to $2.661 \pm 0.037~\%$ in terms of the full width at half maximum (FWHM).

These results show the potential of SPCM for applications in neutron beam benchmarking and spectroscopy.
By combining SPCM with the time-of-flight (TOF) method, both the monochromaticity and velocity of a neutron beam can be evaluated.
The applicable energy range of an SPCM is determined by the performance of its polarizer and analyzer such as polarizing supermirrors and He-3 spin filters\cite{hino2006development, okudaira2020development}. 
Operating with wider energy range spin polarizing devices enables SPCM operation over broader energy range.
Furthermore, the combination of SPCM and TOF techniques provides a basis for new inelastic scattering spectrometers,
in which the TOF method defines the incident energy, while SPCM determines the scattered neutron energy. 
Owing to its high sensitivity to monochromaticity, 
the spectrometer with SPCM provides a high-resolution analyzer for quasi-elastic scattering, 
where detecting minute broadening of the energy spectrum is crucial. 

\section{Summary}
We have presented spin phase continuous modulation (SPCM) and demonstrated its capability for quantitative evaluation of neutron velocity and monochromaticity based on theoretical formulation and experimental verification.
The formulation shows that the phase and phase dispersion of the SPCM signal are determined by the neutron velocity and monochromaticity, 
respectively, and scale proportionally with the distance between the modulators and the modulation frequency.
This relationship provides a predictable basis for characterizing neutron beam properties.
To validate the formulation, experiments were performed using three modulation frequencies and five distances between the RF coils.
The measured phase and phase dispersion showed reasonable agreement with the theoretical prediction.
From the obtained phase dispersion, the monochromaticity of the beamline was determined to be $2.661 \pm 0.037~\%$ in terms of the full width at half maximum (FWHM).
The present results suggest that SPCM has potential applications in neutron beam benchmarking and spectroscopy.
In combination with the time-of-flight (TOF) method and wide-energy-range spin-polarizing devices, SPCM may provide a new approach for neutron beam characterization and spectroscopy over a broad energy range.
Further technical development of SPCM is required to expand its applicability.

\section*{Acknowledgment}
The neutron experiments at JRR-3 C3-1-2-2 were carried out by the JRR-3 general user program managed by the Institute for Solid State Physics,
the University of Tokyo (proposal No. 24583, 24407, 25567).
The polarizing devices fabrication work has been conducted under the visiting researcher’s program of the Institute for Integrated Radiation and Nuclear Science, Kyoto University.
The development of polarizing neutron mirrors was also supported by the visiting Researcher’s Program of the Research Reactor Institute, Kyoto University.
The financial supports come from JSPS KAKENHI (Grant No. 23K23274) and JST FOREST Program (Grant No.JPMJFR2237).

\bibliography{bibliography}

\end{document}